\documentclass[12pt]{iopart}
\usepackage{iopams}
\usepackage{graphicx}
\usepackage[abbr,dcucite]{harvard}
\usepackage[hidelinks]{hyperref}
\usepackage[mathlines]{lineno}

\begin{document}

\paper[Biological dose representation for carbon-ion radiotherapy]{Biological dose representation for carbon-ion radiotherapy of unconventional fractionation}

\author{Nobuyuki~Kanematsu$^{1,2}$ and Taku~Inaniwa$^2$}

\address{$^1$ Medical Physics Section, National Institute of Radiological Sciences Hospital, 4-9-1 Anagawa, Inage-ku, Chiba 263-8555, Japan}
\address{$^2$ Department of Accelerator and Medical Physics, National Institute of Radiological Sciences, 4-9-1 Anagawa, Inage-ku, Chiba 263-8555, Japan}

\ead{kanematsu.nobuyuki@qst.go.jp} 

\begin{abstract}
In carbon-ion radiotherapy, single-beam delivery each day in alternate directions has been commonly practiced for operational efficiency, taking advantage of the Bragg peak and the relative biological effectiveness (RBE) for uniform dose conformation to a tumor.
The treatment plans are usually evaluated with total RBE-weighted dose, which is however deficient in relevance to the biological effect in the linear-quadratic model due to its quadratic-dose term, or the dose-fractionation effect. 
In this study, we reformulate the extrapolated response dose (ERD), or synonymously BED, which normalizes the dose-fractionation and cell-repopulation effects as well as the RBE of treating radiation, based on inactivation of a single model cell system and a typical treating radiation in carbon-ion RT.
The ERD distribution virtually represents the biological effect of the treatment regardless of radiation modality or fractionation scheme.
We applied the ERD formulation to simplistic model treatments and to a preclinical survey for hypofractionation based on an actual prostate-cancer treatment of carbon-ion radiotherapy. 
The proposed formulation was demonstrated to be practical and to offer theoretical implications.
In the prostate-cancer case, the ERD distribution was very similar to the RBE-weighted-dose distribution of the actual treatment in 12 fractions. 
With hypofractionation, while the RBE-weighted-dose distribution varied significantly, the ERD distribution was nearly invariant, implying that the carbon-ion radiotherapy would be insensitive to fractionation.
However, treatment evaluation with simplistic biological dose is intrinsically limited and must be complemented in practice somehow by clinical experiences and biology experiments.
\end{abstract}

\pacs{87.53.-j, 87.53.Tf, 87.64.-t}

\vspace{2pc}
\noindent{\it Keywords\/}: linear-quadratic model, biologically effective dose, relative biological effectiveness, radiation treatment planning, multimodal radiotherapy


\section{Introduction}

The basis of radiotherapy (RT) for cancer treatment lies in radiobiology of human tissues and cells.
Douglas and Fowler \citeyear{Douglas1976} first proposed a formula for cell-survival fraction as an exponential linear-quadratic (LQ) function of instantly delivered dose, on the hypothesis that cell inactivation by lethal or double-sublethal damage was correlated with biological reactions.
In fractionated RT, the relative effectiveness per dose increases with fraction dose according to the LQ model, which is referred to as the dose-fractionation effect and is generally valid up to 10 Gy fractions \cite{Fowler1989}. 
In clinical practice, radiation oncologists evaluate the total dose accumulated over the fractions for the assessment of treatment.
However, the total doses of different fractionation or of uneven fractionation with dose varying from fraction to fraction cannot be compared directly due to the dose-fractionation effect.
The concept of extrapolated response dose (ERD), which is more commonly referred to as biologically effective dose (BED), was introduced to universally represent the treatment dose by the total dose in infinite fractions for the same effect on cells \cite{Barendsen1982}.
The ERD extrapolates the dose-fractionation effect to its limit for universal assessment of RT treatments.

Besides dose fractionation, radiation quality modifies the relative effectiveness per dose especially of ions whose linear energy transfer (LET) rises with depth and also causes a Bragg peak in dose.
The use of such ion beams for RT was pioneered in the United States with dose prescription incorporating depth-dependent weighting of relative biological effectiveness (RBE) to give a uniform biological effect (BE) in a spread-out Bragg peak (SOBP) \cite{Castro1993}, followed by carbon-ion RT mainly in Japan and Germany. 
The achieved high uniformity of the RBE-weighted dose (RWD) to a tumor facilitates the delivery of single beams with daily different dose distributions for operational efficiency. 
Such uneven dose fractionation except to tumor has been commonly practiced in Japan often involving clinical studies toward hypofractionation \cite{Kamada2015}.
In Germany, carbon-ion beams have been occasionally used as a boost in multimodal RT \cite{Combs2013}. 
Nevertheless, these treatments may still be evaluated unconsciously with the total RWD distribution despite its deficiency in additivity. 

Dale and Jones \citeyear{Dale1999} extended the ERD concept to include RBE for high-LET radiations. While their formulation has continually been reviewed and applied to radiobiological studies \cite{CarabeFernandez2007,Fowler2010,Holloway2013}, its direct use in treatment planning has not been realized in carbon-ion RT. 
One reason may be their principle of formulation with physical dose and radiosensitivities per cell type per radiation type, which are not directly relevant to the treatments in the clinic and are demanding for carbon-ion beams of continuously varying radiation quality.

In this study, we attempt to apply the ERD concept to practical and valid assessment of carbon-ion RT treatment.
In the following sections, we reformulate the ERD from basic radiobiology, apply it to some simplistic model treatments and to a typical prostate-cancer treatment and demonstrate its significance and usability.

\section{Methods and Materials}

\subsection{Radiobiological modeling} \label{sec:2-1}

\paragraph{Survival factor}
In radiobiology, survival \emph{fraction} is defined as a fraction of clonogenic cells surviving a radiation exposure.
In this paper, survival \emph{factor} is termed to additionally include the effect of inter-fraction cancer-cell repopulation.
On the hypothesis of constant rates for cell division and natural loss \cite{Dale1989}, the survival factor at time $t$ after the delivery of $i$-th fraction dose $D_i$ at time $t_i$ is formulated as
\begin{eqnarray} \label{eq:1}
S_i(t) = {\rm e}^{-\alpha_i D_i -\beta_i D_i^2} 2^{(t-t_i)/T_d},
\end{eqnarray}
where $\alpha_i$ and $\beta_i$ are the LQ coefficients and $T_d$ is the effective doubling time for the surviving cancer cells, which we assumed to equal the tumor-doubling time.
The survival factor at the end of a treatment in $n$ fractions is formulated as
\begin{eqnarray} \label{eq:2}
S = \prod_{i=1}^{n-1} S_i(t_{i+1})\, S_n(t_n) = \prod_{i=1}^n {\rm e}^{-\alpha_i D_i - \beta_i D_i^2} \cdot {\rm e}^{ T \ln2/T_{\rm d}},
\end{eqnarray}
where $T = t_n-t_1$ is the overall treatment time.

\paragraph{Biological effect}
The BE for an instant beam delivery of fraction $i$ is defined as
\begin{eqnarray} \label{eq:3}
{E_{\rm B}}_i = -\ln S_i(t_i) = \alpha_i D_i + \beta _i D_i^2,
\end{eqnarray}
which statistically corresponds to the mean number of unrepaired lethal damages per cell.
The BE for an overall treatment is similarly defined as
\begin{eqnarray} \label{eq:4}
E_{\rm B}= -\ln S = \sum_{i=1}^{n} \alpha_i D_i + \sum_{i=1}^{n} \beta _i D_i^2 - \frac{T \ln2}{T_{\rm d}},
\end{eqnarray}
where the last term represents the cell-repopulation effect.

\paragraph{RBE-weighted dose}
The RWD is defined as the dose of a reference radiation with LQ coefficients $\alpha_{\rm ref}$ and $\beta_{\rm ref}$ to cause the same BE,
\begin{eqnarray} \label{eq:5}
{E_{\rm B}}_i = \alpha_i D_i + \beta_i D_i^2 = \alpha_{\rm ref} {D_{\rm RW}}_i + \beta_{\rm ref} {D_{\rm RW}}_i^2,
\end{eqnarray}
with symmetric solution
\begin{eqnarray} \label{eq:6}
D_{\{ i,\:{\rm RW} i\}} = \frac{\alpha_{\{i,\:{\rm ref}\}}}{\beta_{\{i,\:{\rm ref}\}}} \left( \sqrt{ \frac{1}{4} + \frac{\beta_{\{i,\:{\rm ref}\}}}{\alpha_{\{i,\:{\rm ref}\}}^2} {E_{\rm B}}_i} - \frac{1}{2} \right),
\end{eqnarray}
which also define the RBE of the interest radiation and the total RWD,
\begin{eqnarray} \label{eq:7}
\epsilon_i = \frac{{D_{\rm RW}}_i}{D_i} \quad \mbox{and} \quad
D_{\rm RW} = \sum_{i=1}^{n} {D_{\rm RW}}_i = \sum_{i=1}^{n} \epsilon_i\, D_i .
\end{eqnarray}
The total RWD is widely used for carbon-ion RT treatment, though deficient against dose fractionation due to quadratic-term contribution and against cancer-cell repopulation between fractions.

\paragraph{Extrapolated response dose}
In accordance with Dale and Jones \citeyear{Dale1999}, the ERD is defined as a total physical dose of reference radiation in a hypothetical treatment of infinite fractionation for the same BE as with the actual treatment.
According to Barendsen \citeyear{Barendsen1982} and Dale \citeyear{Dale1989}, the ERD is the ratio of the BE to the $\alpha$ parameter of interest, which is for the reference radiation in this case, or 
\begin{eqnarray} \label{eq:8}
{D_{\rm ER}}_i = \frac{{E_{\rm B}}_i}{\alpha_{\rm ref}} = {D_{\rm RW}}_i + \frac{\beta_{\rm ref}}{\alpha_{\rm ref}} {D_{\rm RW}}_i^2,
\end{eqnarray}
which consists of the linear and quadratic dose terms. 
The ERD for an overall treatment is similarly defined as
\begin{eqnarray} \label{eq:9}
D_{\rm ER} = \frac{E_{\rm B}}{\alpha_{\rm ref}} = D_{\rm RW} + \frac{\beta_{\rm ref}}{\alpha_{\rm ref}} \sum_{i=1}^{n} {{D_{\rm RW}}_i}^2 - \frac{1}{ \alpha_{\rm ref}} \frac{T \ln2}{T_{\rm d}},
\end{eqnarray}
which additionally includes the cell-repopulation term.
This RWD-based formulation is mathematically equivalent to the original formulation by Dale and Jones \citeyear{Dale1999} and easier to use in clinical practice of carbon-ion RT, where the RWD is available.

\paragraph{Tumor-control probability}
The tumor-control probability (TCP) is radiobiologically modeled as the probability of inactivating all of the $N_0$ cancer cells originally existed in a tumor \cite{Munro1961} and is statistically given by 
\begin{eqnarray} \label{eq:10}
P_{\rm TC} = {\rm e}^{-N_0 S} = \exp\left(-N_0 {\rm e}^{-E_{\rm B}}\right),
\end{eqnarray}
which should be high for a curative treatment. 
Inversely, when the number of cancer cells is reasonably estimated, a curative TCP can be translated into treatment BE
\begin{eqnarray} \label{eq:11}
E_{\rm B} = \ln\frac{N_0}{-\ln P_{\rm TC}}.
\end{eqnarray}
With a compensation for cancer-cell repopulation between fractions, the treatment BE may be evenly divided into $n$ fractions of instant BE
\begin{eqnarray} \label{eq:12}
{E_{\rm B}}_1 = \frac{1}{n} \left( \ln\frac{N_0}{-\ln P_{\rm TC}} + \frac{T \ln2}{T_{\rm d}} \right),
\end{eqnarray}
from which fraction dose $D_1$ and RWD ${D_{\rm RW}}_1$ can be determined by \eref{eq:6} to prescribe optimum beam deliveries for curative RT.

\paragraph{Beam delivery}
A treatment is normally prescribed with RWD to a tumor, which is inversely converted with RBE to a physical dose to a reference point for beam-delivery control or assessment.
When multiple beams ($b$) are involved in a fraction, the RBE for the fraction dose is calculated with the dose-mean LQ coefficients,
\begin{eqnarray} \label{eq:13}
D_i = \sum_b D_{i_b}, \quad
\alpha_i = \sum_b \alpha_{i_b} \frac{D_{i_b}}{D_i} \quad \mbox{and} \quad
\sqrt{\beta_i} = \sum_b \sqrt{\beta_{i_b}} \frac{D_{i_b}}{D_i}
\end{eqnarray}
for the mixed radiation in the LQ model \cite{Zaider1980}.
To allot the prescribed RWD to the relevant beams as specified, the physical beam doses are generally derived iteratively by the LQ model or deterministically by the lesion-additivity (LA) model \cite{Lam1987} as approximation. 

\subsection{Application to simplistic model treatments}

We investigated these representations of treatment dose in simplistic examples with four model radiations of two modalities: photon radiation and three carbon-ion radiations sampled in a typical SOBP with respective $\alpha =$ 0.3, 0.5, 1.0 and 1.5 Gy$^{-1}$ and a common $\beta =$ 0.06 Gy$^{-2}$ for radiosensitivity of $N_0 = 10^7$ hypothetical cancer cells  in a hypothetical fast-growing tumor with doubling time $T_{\rm d} = 30$ days, which roughly mimicked a realistic situation \cite{Inaniwa2015}.  
In the following model treatments, a prescribed fraction RWD taking the photon radiation for a reference was assumed to be delivered once a day, seven days a week for simplicity. 
We generally note prescribed RWD values with postfix ``(RBE)'' to clarify that they are RBE-weighted.

\paragraph{Dose fractionation} 
For a carbon-ion RT treatment of total 40 Gy (RBE), we varied the number of fractions to prescribe evenly.

\paragraph{Multimodal RT}
For a treatment initially with photons of 2 Gy per fraction for 10 days, followed by carbon ions of 4 Gy (RBE) per fraction for 6 days to total 44 Gy (RBE) in 16 days, we evaluated the accumulation of treatment dose.

\subsection{Application to a prostate-cancer treatment}

\paragraph{Clinical dosimetry system}
At the National Institute of Radiological Sciences (NIRS) in Japan, carbon-ion RT doses are prescribed in clinical dose, defined as 
\begin{eqnarray} \label{eq:14}
{D_{\rm C}}_i = f_{\rm C}\, {D_{\rm RW}}_i = f_{\rm C}\, \epsilon_i\, D_i ,
\end{eqnarray}
where clinical factor $f_{\rm C} = 2.41$ was introduced for historical reasons and RBE $\epsilon_i$ was defined against a typical carbon-ion beam at a central SOBP depth as a reference radiation on the inactivation of incubative human salivary gland (HSG) tumor cells as an endpoint, resulting in the LQ coefficients of $\alpha_{\rm ref} = 0.764\ {\rm Gy}^{-1}$ and $\beta_{\rm ref} = 0.0615\  {\rm Gy}^{-2}$ \cite{Inaniwa2015}. 
This is deviating from the conventional RBE defined against photon radiation with the LQ coefficients of $\alpha_{\rm x} = 0.313$ Gy$^{-1}$ and $\beta_{\rm x} = 0.0615$ Gy$^{-2}$ for the same endpoint \cite{Furusawa2000}.
The RWD is a biologically equivalent dose of the reference carbon-ion radiation and the clinical dose further involves artificial rescaling to it. 
For distinction, we note clinical-dose values with postfix ``(C)'' so that 1 Gy (RBE) corresponds to 2.41 Gy (C).

\paragraph{Treatment}
For demonstration, we took a case of prostate-cancer patient who received carbon-ion RT in 12 fractions of 4.3 Gy (C) over 3 weeks \cite{Nomiya2014}. 
In the planning CT of the patient immobilized in a supine position, the clinical target volume (CTV) included the prostate and the seminal vesicles. 
The planning target volume (PTV) additionally included anterior and lateral margins of 10 mm and a posterior margin of 5 mm.
Lateral opposing carbon-ion beams were used alternately for the initial 8 fractions to cover the original PTV (PTV1) with more than 95\% of the prescribed fraction dose, or cumulatively with about 2/3 of the prescribed total dose.
To care against the risk of complication, the rectum was fully excluded from the PTV to derive a restricted PTV (PTV2), for which similar opposing beams with shrunk fields were used alternately for the remaining 4 fractions. 
The daily beam delivery was conducted with a pencil-beam scanning technique \cite{Furukawa2010} to conform 4.3 Gy (C) to the PTV with a single beam per fraction. 
The physical and clinical dose distributions per fraction were calculated and stored in the treatment plan while the total clinical dose distribution was primarily used for clinical plan assessment.

\paragraph{Plan distributions}
On the assumption of quadratic-coefficient invariance among radiations \cite{Inaniwa2015} in \eref{eq:5}, the $\alpha/\beta$-ratio for each fraction was obtained from a set of RWD ${D_{\rm RW}}_i = {D_{\rm C}}_i /f_{\rm C}$ and physical dose $D_i$ as
\begin{eqnarray} \label{eq:15}
\frac{\alpha_i}{\beta_{\rm ref}} = \frac{{D_{\rm RW}}_i}{D_i} \left( \frac{\alpha_{\rm ref}}{\beta_{\rm ref}} + {D_{\rm RW}}_i - D_i \right).
\end{eqnarray}
Using these data originated from the plan, we calculated the distributions of total physical dose by $\sum_i D_i$, total dose-mean $\alpha/\beta$ ratio by $\sum_i (\alpha_i/\beta_{\rm ref}) D_i /\sum_i D_i$, total clinical dose by $\sum_i {D_{\rm C}}_i$ and ERD by \eref{eq:9}, where we ignored the cell-repopulation effect on the assumption of slow-growing prostate cancer with $T \ll T_{\rm d}$.

\paragraph{Hypofractionation}
Aside from the actual treatment, we attempted a survey toward hypofractionation, where we reduced number of beams from four to two to simplify the fractionation scheme while conserving the total dose.
To simulate scanning beams of stepped target dose, we fused the original-filed beams by 2/3 and the shrunk-field beams by 1/3 per direction into the left and right beams and obtained their physical-dose and $\alpha/\beta$-ratio distributions.
In addition, to simulate even fractionation with the left and right beams delivered successively each day, we further fused them by 1/2 each and obtained the physical-dose and $\alpha/\beta$-ratio distributions of a fraction.
We virtually varied number of fractions for each of the successive and alternate delivery schemes.
For the same ERD of 24.49 Gy to the prostate as with 12 fractions of 4.3 Gy (C) by \eref{eq:9}, we additionally prescribed fraction clinical doses of 6.124, 10.83 and 18.31 Gy (C) by \eref{eq:6} with ${E_{\rm B}}_i = ({D_{\rm ER}}_i/\alpha_{\rm ref})/n$ for $n=$ 8, 4 and 2 fractions, respectively.
Accordingly, we rescaled the respective fraction physical-dose distributions by the same factors as for the prescribed clinical doses, or by 1.424, 2.519 and 4.258, based on the fact that the reference radiation quality with an invariant RBE of 1 was in the prostate somewhere that was implicitly taken as a dose reference point.
We then obtained the ERD distribution for each $n$ from the fraction physical-dose and $\alpha/\beta$-ratio distributions using \eref{eq:5}--\eref{eq:9}.

\section{Results}

\subsection{Application to simplistic model treatments}

\begin{figure}
\begin{indented}
\item[] \includegraphics[width=8.5cm]{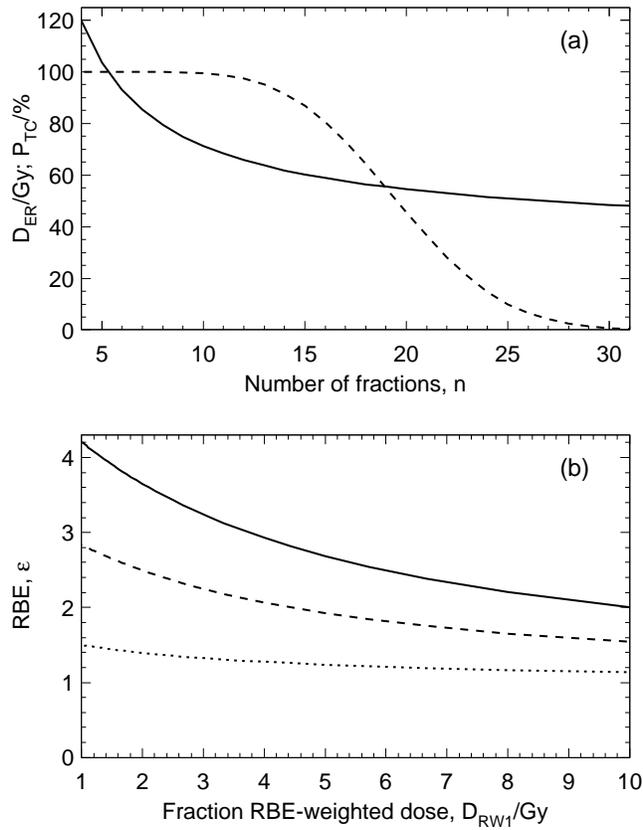}
\end{indented}
\caption{\label{fig:1} Dose fractionation effect for evenly fractionated carbon-ion RT of total 40 Gy (RBE): 
(a) ERD ($D_{\rm ER}$, \full) and TCP ($P_{\rm TC}$, \dashed) as functions of number of fractions.
(b) RBEs for low-$\alpha$ (\dotted), mid-$\alpha$ (\dashed) and high-$\alpha$ (\full) carbon-ion radiations as functions of fraction RWD. }
\end{figure}

\paragraph{Dose fractionation} 
\Fref{fig:1}(a) shows the representations of treatment dose for a prescription of total 40 Gy (RBE). 
The ERD decreased with number of fractions due to dose fractionation and cancer-cell repopulation and correlated with the TCP.
\Fref{fig:1}(b) shows the RBEs of the three carbon-ion radiations in a SOBP.
The variation of the RBEs as well as their values decreased with the fraction RWD, which implies that the SOBP must be designed differently for uniform treatment according to the prescribed fraction RWD.

\begin{figure}
\begin{indented}
\item[] \includegraphics[width=8.5cm]{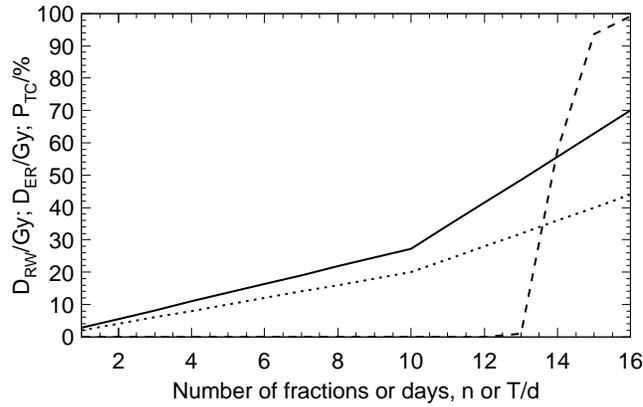}
\end{indented}
\caption{\label{fig:2}
A multimodal RT treatment in ten fractions of 2 Gy with photons and six fractions of 4 Gy (RBE) with carbon ions over 16 days: cumulative RWD ($D_{\rm RW}$, \dotted), ERD ($D_{\rm ER}$, \full) and TCP ($P_{\rm TC}$, \dashed) as functions of number of fractions or days.}
\end{figure}

\paragraph{Multimodal RT}
\Fref{fig:2} shows the representations of daily cumulative dose for the model multimodal RT treatment.
By the change of fraction dose from 2 Gy with photons to 4 Gy (RBE) with carbon ions after day 10, the ERD slope changed by a larger factor of 2.62 due to the quadratic term.
On day 16 in the end of the treatment, the cell-repopulation term reduced the ERD by 1.16 Gy or 1.7\% of the total ERD, while the TCP of 99.2\% may be reasonably curative. 
If the same total RWD of 44 Gy (RBE) were evenly delivered in 16 fractions of 2.75 Gy (RBE), the TCP would be reduced to 98.1\% according to the formulas in \sref{sec:2-1}, indicating deficiency of the RWD-based prescription against the change of dose fractionation.
To obtain the same TCP of 99.2\% with even 16 fractions, the required RWD would be 2.83 Gy (RBE) per fraction or 45.3 Gy (RBE) in total.

\subsection{Application to a prostate-cancer treatment}

\begin{figure}
\begin{indented}
\item[] \includegraphics[width=8.5cm]{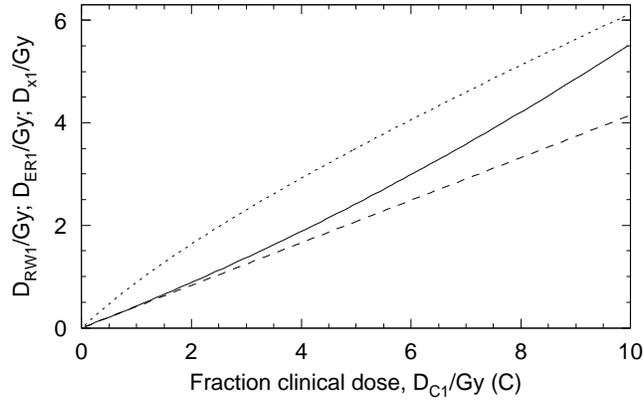}
\end{indented}
\caption{\label{fig:3} Relation between fraction clinical dose (${D_{\rm C}}_1$) and fraction RWD (${D_{\rm RW}}_1$, \dashed), instant ERD (${D_{\rm ER}}_1$, \full) and the photon-equivalent dose (${D_{\rm x}}_1$, \dotted) in the NIRS clinical dosimetry system.}
\end{figure}

\Fref{fig:3} shows the relation between fraction clinical dose and fraction RWD, instant ERD and the photon-equivalent dose in the NIRS clinical dosimetry system. 
The approximation between ERD and RWD at small fraction sizes is due to minor contribution of the quadratic term, or ${D_{\rm RW}}_1 \ll \alpha_{\rm ref}/\beta_{\rm ref}$ in \eref{eq:8}.

\begin{figure}
\begin{indented}
\item\includegraphics[width=13cm]{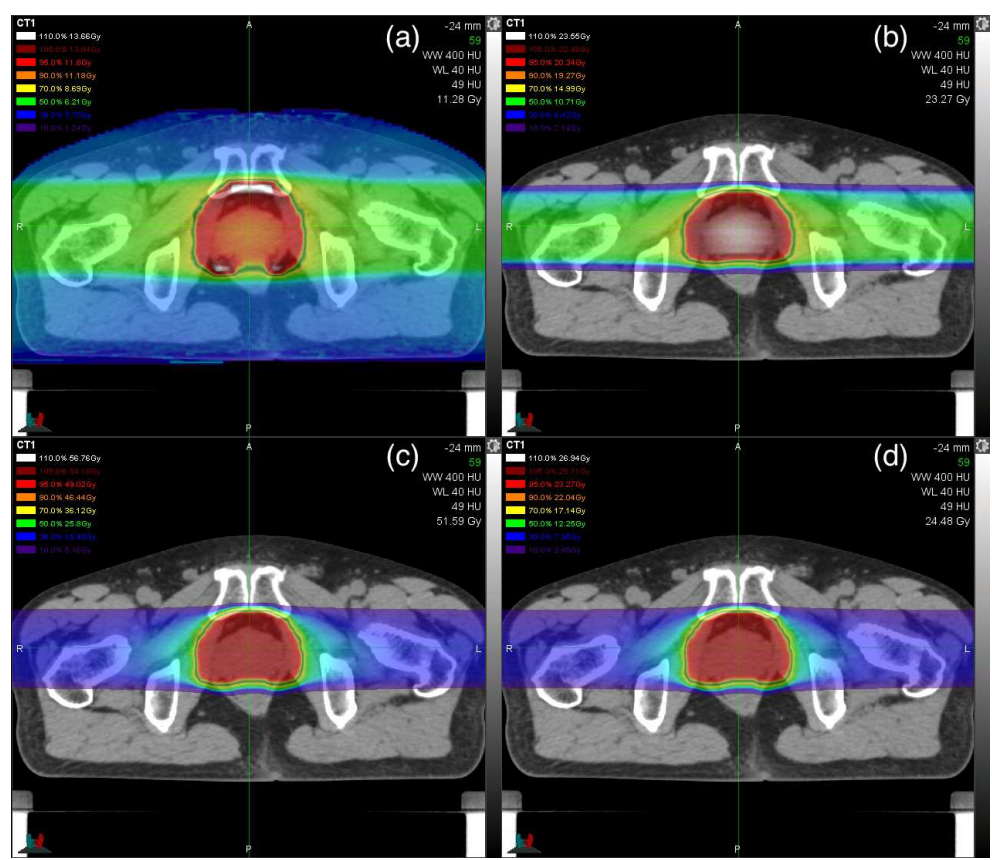}
\end{indented}
\caption{\label{fig:4} Planning CT image of the prostate-cancer patient in the isocenter plane with green crosshairs for the right--left and anterior--posterior axes and overlaid color wash (10\%, 30\%, 50\%, 70\%, 90\%, 95\%, 105\% and 110\%): (a) dose-mean $\alpha/\beta$ ratio relative to 12.42 Gy, (b) total physical dose relative to 21.41 Gy, (c) total RWD relative to 51.6 Gy (RBE) and (d) ERD relative to 24.49 Gy.}
\end{figure}

\begin{figure}
\begin{indented}
\item[] \includegraphics[width=13cm]{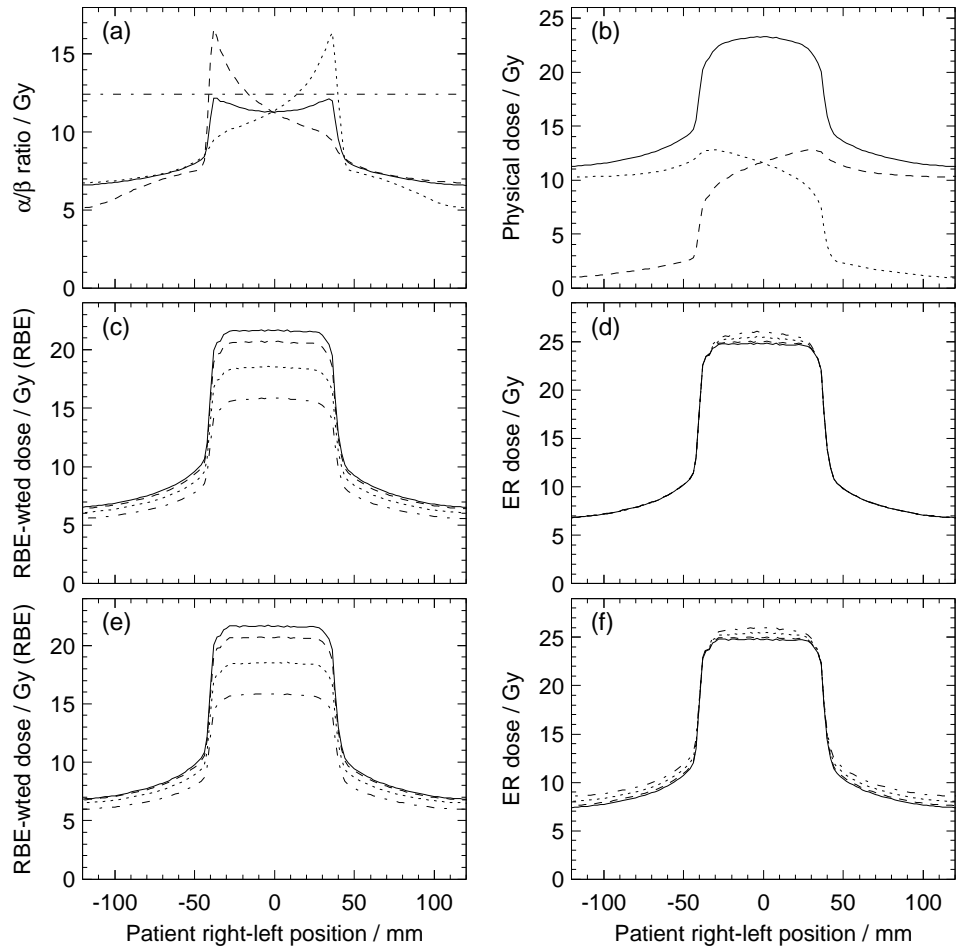}
\end{indented}
\caption{\label{fig:5} Profiles on the patient right--left axis for the prostate-cancer treatment plans: 
(a) dose-mean $\alpha/\beta$ ratio for inactivation of incubative HSG tumor cells, (b) total physical dose for 12 fractions, (c) total RWD and (d) ERD with successive beam delivery, and (e) total RWD and (f) ERD with alternate beam delivery. 
In (a) and (b) are drawn the contributions of the left (\dashed) and right (\dotted) beams and their total (\full) with the $\alpha_{\rm ref}/\beta_{\rm ref}$ = 12.42 Gy level (\chain).
In (c)--(f) are drawn the treatment-plan doses for 12 (\full), 8 (\dashed), 4 (\dotted) and 2 (\chain) fractions prescribed for 24.49 Gy in ERD.}
\end{figure}

\Fref{fig:4} shows the dose distributions calculated for the actual prostate-cancer treatment.
The dose-mean $\alpha/\beta$-ratio and physical-dose distributions were gentle in the opposing beam arrangement while the $\alpha/\beta$ ratio was high in the anterior and posterior sides of the prostate and the physical dose was high in the central prostate.
The relative difference between the clinical dose in \fref{fig:5}(c) and the ERD in \fref{fig:5}(d) was minor due to small quadratic-term contribution at the level of 4.3 Gy (C) or 1.78 Gy (RBE) as consistent with \fref{fig:3}.

Figures \ref{fig:5}(a) and \ref{fig:5}(b) show the profiles of total physical dose and dose-mean $\alpha/\beta$ ratio, where the two opposing beams were designed for 12 fractions of 4.3 Gy (C) and were also reused for the hypofractionated treatment plans with rescaling to conserve the ERD to the point with $\alpha/\beta =\alpha_{\rm ref}/\beta_{\rm ref}$.
The RWD distribution in \fref{fig:5}(c) was severely deformed with hypofractionaton, while the ERD distribution in \fref{fig:5}(d) was nearly invariant, so that the plans of different fractionation cannot be compared with RWD even relatively.
The degradation of the SOBP with hypofractionation was apparently caused by the forced reuse of the nonopimal beams in this analysis. 
In the outside of the SOBP in figures \ref{fig:5}(e) and \ref{fig:5}(f), the alternate beam delivery increased the RWD and ERD themselves and the influence of hypofractionation on ERD.

\section{Discussion}

Biological dosimetry is a concept of radiation dose measurement by consequential response of a reference biological system.
The NIRS clinical dose and its relevant RWD and ERD are model-based doses that offer virtual {\it in vivo} biological dosimetry with an incubative HSG tumor cell of moderate radiosensitivity \cite{Furusawa2000} as a reference model cell system, whose choice is left as a major arbitrariness beyond this study.
This simplification is distinctive among ERD formulations for high-LET radiation since Dale and Jones \citeyear{Dale1999} and enabled application to patient dose distributions.
The ERD theoretically indicates a universal dose that is directly related to the BE on the reference cells, which may be useful for preclinical and retrospective studies involving various radiation modalities or dose fractionations.
If the initial cancer-cell population-density distribution $\rho_0(\vec{r}) = {\rm d}^3 N_0/{\rm d}\vec{r}$ as well as the ERD distribution is somehow known, the TCP in \eref{eq:10} is modified to 
\begin{eqnarray} \label{eq:16}
P_{\rm TC}= \exp\left( - \int\!\!\!\!\! \int\!\!\!\!\! \int \rho_0(\vec{r}) \, 
 {\rm e}^{- \alpha_{\rm ref}\,D_{\rm ER}(\vec{r})} \, {\rm d}\vec{r} \right) 
\end{eqnarray}
for dose-distribution assessment under the hypothesis that the tumor is formed by the incubative reference cells.

In reality, the actual cancer and normal cells in a patient may be substantially different from the reference in intrinsic radiosensitivity and in environmental conditions including oxygen and biochemical concentrations, various cell interactions in tissues, etc.
The clinical sensitivity to the cell response may also vary among diseases and individuals.
There may also be differences in the time structure of radiation exposure between the treatment and the biology experiment that determined the reference radiosensitivity \cite{Inaniwa2013}. 
As a result, these biological doses may not directly be relevant to the prognosis of treatment.
Another major arbitrariness exists in the reference radiation, which should be chosen to minimize the overall inaccuracy of cancer treatment.
For example, Kanematsu \etal \citeyear{Kanematsu2002} found a few percent inconsistency between the LQ and LA models for RBE of a mixed carbon-ion radiation defined against a photon radiation, which should have been minimized if the RBE had been defined to only relate similar-radiation doses.
The NIRS clinical dosimetry system, which is based on the RBE defined against a typical treating radiation, is therefore advantageous for accurate prescription of tumor doses in the modality-specific scale without potentially inaccurate translation to conventional-RT doses.
In fact, carbon-ion RT has been conducted according to disease-specific treatment protocols with abundant clinical experiences including dose-escalation studies \cite{Kamada2015}.
The clinically determined curative doses per disease and organ tolerance doses should thus be reflecting all the differences between the reference biology experiment and the actual cancer treatments of carbon-ion RT in its own dose scale.

In the prostate cancer treatment, the ERD distribution was very similar to the clinical-dose distribution with 12 fractions of 4.3 Gy (C), which happens to be a typical practice of carbon-ion RT \cite{Kamada2015}.
In such a case, the clinical dose distribution can be approximately interpreted as the ERD distribution with rescaling by
\begin{eqnarray} \label{eq:17}
\frac{D_{\rm ER}}{D_{\rm C}} = \frac{1}{f_{\rm C}} \left( 1+\frac{\beta_{\rm ref}}{\alpha_{\rm ref}} \frac{\check{D}_{\rm C}{}_1}{ f_{\rm C}} \right),
\end{eqnarray}
where $\check{D}_{\rm C}{}_1$ is the fraction clinical dose prescribed to a tumor.
The hypofractionation attempted for prostate cancer treatment apparently degraded the dose concentration of the total RWD distribution, which was inconsistent with the ERD distribution and thus may indicate general deficiency of RWD for hypofractionated treatments.
The observed invariance of ERD may have been caused by accidental cancelation between the SOBP (high dose, high $\alpha/\beta$) and its outside regions (low dose, low $\alpha/\beta$) in the relative dose-fractionation effect $(1 + D_i\,\beta_i/\alpha_i)$.
The cancellation may generally be valid for carbon-ion beams because their physical dose and LET are naturally correlated, suggesting that carbon-ion RT may generally be insensitive to fractionation.
In reality, however, differentiation between cancer and normal cells in radiosensitivity is necessary to evaluate the therapeutic gain by fractionation \cite{Yoshida2015}, or could be learned from clinical experiences retrospectively \cite{Fukahori2016}.

The use of uneven fractionation such as with alternate single-beam delivery may be a matter of clinical decision to balance between operational efficiency and dose concentration to a tumor, which is not normally very sensitive if adequately evaluated with ERD. 
The field-shrinking approach taken for the actual prostate-cancer treatment, originated from the limitations with historical passive broad-beam delivery, should be replaced by the field-modulation approach for multiple target volumes and doses with pencil-beam scanning, which will simply improve the efficiency by reducing the number of beams to plan and to verify for quality assurance. 

\section{Conclusions}

The ERD, or synonymously BED, is a representation of treatment dose that normalizes the effects of dose fractionation, inherent tumor growth and the RBE of treating radiation.
For assessment of carbon-ion RT treatment, we simplified the ERD concept to biological dose for a single model cell system and reformulated it as a derivative of clinical dose used in the clinic.
The ERD will theoretically be useful for preclinical and retrospective studies when variation in fractionation is involved.
For a prostate-cancer treatment of carbon-ion RT, we found that the ERD distribution was very similar to the clinical dose distribution at a normal fraction size, that the ERD distribution was nearly invariant against fractionation, that the clinical dose would not suffice with hypofractionation and that uneven fractionation only slightly degraded the dose concentration of even fractionation.
However, treatment evaluation with simplistic biological dose is intrinsically limited and must be complemented in practice somehow by clinical experiences and biology experiments.

\ack

The authors thank Dr H Tsuji for helpful consultation on the prostate-cancer treatment and the team of treatment-planning dosimetrists, especially W Furuichi for technical assistance and M Wakaisami for team management.

\section*{References}

\bibliographystyle{dcu}
\bibliography{references}

\end{document}